\RequirePackage{lineno}
\documentclass[epj,twocolumn, 12pt]{webofc}
\usepackage[varg]{txfonts}   
\wocname{EPJ Web of Conferences}
\woctitle{XLIV International Symposium on Multiparticle Dynamics 2014}

\DeclareGraphicsExtensions{.png}

\usepackage{amsmath}
\usepackage{comment}
\usepackage{geometry}

\newcommand{\tch}{T_{ch}}
\newcommand{\mub}{\mu_{B}}
\newcommand{\muq}{\mu_{Q}}
\newcommand{\mus}{\mu_{S}}
\newcommand{\gs}{\gamma_{S}}

\newcommand{\snn}{\sqrt{s_{NN}}}

\usepackage[colorlinks=true, pdfstartview=FitH, linkcolor=blue,
citecolor=blue, urlcolor=blue, bookmarksopen=true]{hyperref}
\begin{document}
\title{Identified particle production and freeze-out properties in heavy-ion \\collisions at RHIC Beam Energy Scan program}

\author{Sabita Das~(for the STAR collaboration)\inst{1}\fnsep\thanks{\email{sabita@rcf.rhic.bnl.gov}}
}

\institute{Institute of Physics, Bhubaneswar-751005, India}

\abstract{%
The first phase of Beam Energy Scan~(BES) program at the
Relativistic Heavy-Ion Collider~(RHIC) was 
started in the year 2010 with the aim to study the
several aspects of the quantum chromodynamics~(QCD) phase diagram. The Solenoidal Tracker
At RHIC~(STAR) detector has taken data at $\snn = $ 7.7,
11.5, 19.6, 27, and 39 GeV in Au+Au collisions in the years 2010 and
2011 as part of the BES programme. For these beam energies, we
present the results on the particle
yields, average transverse mass and particle ratios for identified
particles in mid-rapidity~($|y|$ < 0.1). The measured particle ratios have been used to study the chemical freeze-out
dynamics within the framework of a statistical model.
}
\maketitle
\section{Introduction}
\label{intro}
To understand the properties of matter under extreme conditions of
high temperature or density, heavy-ion collision experiments
are conducted at RHIC in BNL and the LHC in CERN. These are the
conditions, in which the deconfined phase of QCD matter, the Quark-Gluon Plasma~(QGP),
is created. It is conjectured that the formed hot and dense partonic
matter rapidly expands and cools down. During the evolution 
it undergoes a transition back to the hadronic
matter~\cite{qgp1, qgp2}. Both RHIC and LHC have confirmed the formation of the
QGP in central Au+Au and Pb+Pb collisions~\cite{qgprhic, qgplhc}. 
In QCD, there are three conserved charges, baryon number $B$,
electric charge $Q$ and strangeness $S$. Thus the equilibrium thermodynamic state of
QCD matter is completely determined by temperature $\tch$ and the three chemical
potentials $\mub, ~\muq$, and $\mus$ corresponding to $B$, $Q$ and $S$
respectively. The QCD phase diagram is plotted with the temperature ($T$) as a
function of baryon chemical potential ($ \mu_{B}$)~\cite{tmubth}.
From finite-temperature QCD calculations on the lattice it is theoretically established
that the transition from QGP to a hadron gas happens at high temperature and $\mub$ close to zero
 and is a cross-over~\cite{nature6}. Several QCD-based
 calculations~\cite{1st} suggest existence of first-order phase
transition at a lower $T$ and large $\mu_{B}$. Therefore, there should be an
end point for the first-oder phase transition in the QCD phase
diagram, known as the critical point. Several QCD based models and
also calculations on lattice predict the existence of the critical point at high $\mub$~\cite{qcd1}
and its exact location depends on the
\vspace{0.1cm}
different model assumptions~\cite{qcd2, lqcd1, lqcd2, lqcd3}. It is
worth to mention that not all QCD-based models or lattice groups do
predict the existence of critical point~\cite{nocricpoint}.\\
Theoretically, the phase diagram is explored through non-perturbative QCD
calculations on lattice which indicates the energy scale can be
explored experimentally. Now to explore various
aspects of the QCD phase diagram\cite{qcdphasedgm} such as the search
for the signals of phase boundary, and the search the location of the critical point has become
one of the main goals of the BES program at RHIC. The RHIC BES program
started a test run of Au+Au collisions at $\snn = $ 9.2
GeV~\cite{star9gev}. This was followed by setting a number of observables for the physics
goals~\cite{starnote}. The first phase of the BES program started
in the year in the year 2010 and
2011 in Au+Au collisions at $\snn = $ 7.7, 11.5, 19.6, 27, and 39
GeV. In addition, the NA49 collaboration has reported
evidence that the onset of de-confinement occurs close to $\snn = $ 7.7
GeV~\cite{na49}. The  process of hadron decoupling from an
interacting system in heavy-ion collisions is known as
freeze-out. They are of two types, kinetic and
chemical freeze-out. We will present here the study on chemical
freeze-out, characterised by temperature (T$_{ch}$) and baryon chemical potential ($\mub$), when inelastic
collisions cease and the
particle yields become fixed. The $\tch$ and $\mub$ can be extracted using
the particle ratios obtained from the measured particle yields and
comparing with
model calculations which assume the
system is in chemical and thermal equilibrium.\\\\
\begin{figure*}[!hbtb]
\begin{center}
\includegraphics[scale=0.41,height = 8.3cm,width =8.4cm]{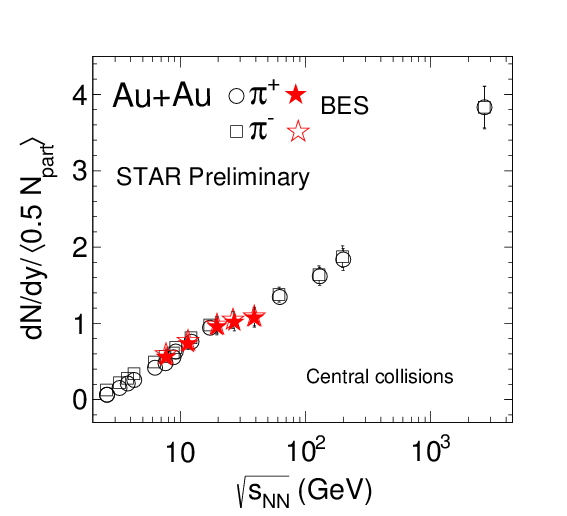}
\includegraphics[scale=0.42,height = 8.3cm,width =8.5cm]{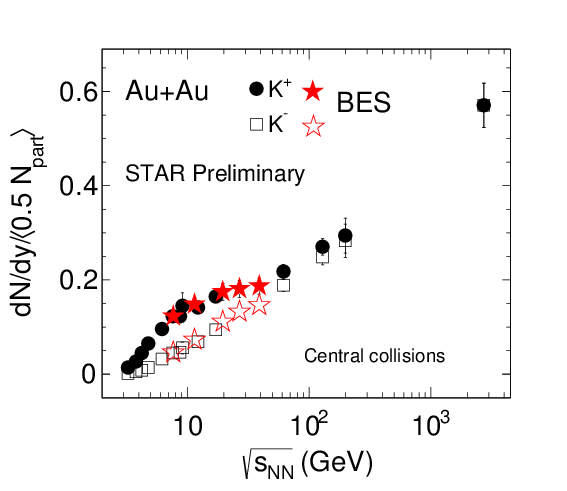}
\includegraphics[scale=0.47]{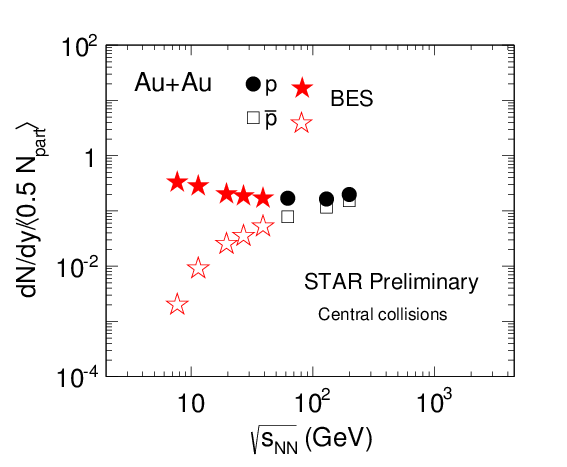}
\caption{$dN/dy$ of $\pi^{\pm}$, $\emph{K}^{\pm}$, $\emph{p}$, and
$\bar{\emph{p}}$ scaled by $\langle 0.5N_{part} \rangle$ as a
function of center of mass energy~($\sqrt{s_{NN}}$) including 
collisions at BES energies~(red points) along with AGS~\cite{dataags1,dataags2,dataags3},
SPS~\cite{datasps}, top RHIC~\cite{datastar},
  and LHC energies~\cite{dataalice} for central
collisions. Errors are statistical
  and systematic errors added in quadrature.}
\label{fEngdndy}
\end{center}
\end{figure*}
Here we will discuss the identified particle production produced
in Au+Au collisions at BES center-of-mass energies $\sqrt {s_{NN} }=
7.7$, 11.5, 19.6, 27, and 39 GeV~\cite{sd, lk}. The energy and centrality
dependence of particle yields, ratios of particle yields, and average transverse
mass will also be discussed for the above energies. 
The mid-rapidity yields of charged pions ($\pi^{\pm}$),
charged kaons ($\emph{K}^{\pm}$), protons ($\emph{p}$,
$\bar{\emph{p}}$),  $K^{0}_{S}$, Lambdas ($\Lambda$, $\bar{\Lambda}$)
and Cascades ($\Xi^{-}$, $\bar{\Xi}^{+}$) have been used for the
chemical freeze-out study~\cite{sd, lk, zhu}. An equilibrium thermal
model, such as THERMUS~\cite{thms} has been quite successful at reproducing the particle
production in heavy-ion collisions~~\cite{cfo1, cfo2, cfo3}. To
extract the chemical freeze-out temperature ($T_{ch}$),
baryon chemical potential ($ \mu_{B}$), strangeness chemical potential
($ \mu_{S}$) and strangeness saturation factor ($\gamma_{S}$),
the experimental particle ratios are used in both grand
canonical ensemble (GCE) and strangeness
canonical ensemble (SCE) approach of the model. 
The centrality and energy dependence of $\tch,
~\mub, ~\mus, ~\gs$ in Au+Au collisions at the above BES energies are studied. 
\begin{figure*}[!hbtb]
\begin{center}
\includegraphics[scale=0.45]{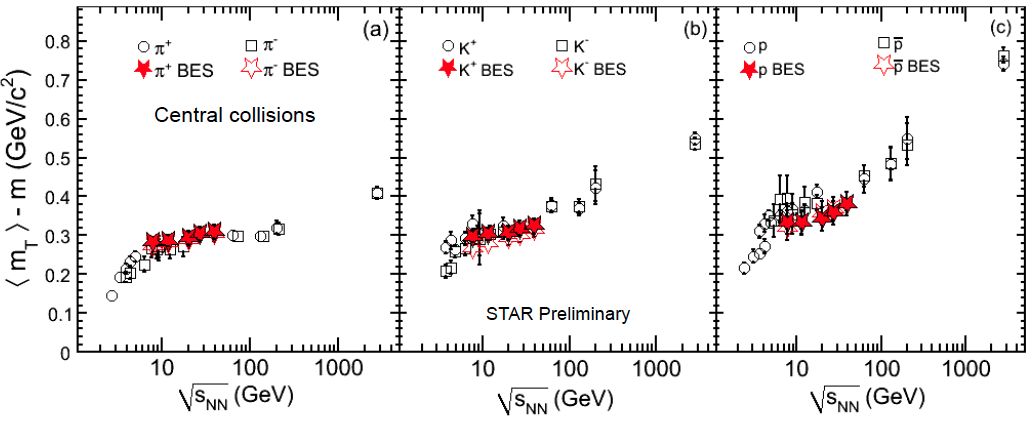}
\caption{The variation of $\langle m_{T} \rangle -m $ of charged pions, kaons, and
  (anti)protons vs. $\sqrt{s_{NN}}$ measured in $|y| <$ 0.1 
in collisions at BES energies along with AGS~\cite{dataags1,dataags2,dataags3},
SPS~\cite{datasps}, RHIC~\cite{datastar},
  and LHC energies~\cite{dataalice} for central collisions. Errors are statistical
  and systematic errors added in quadrature.}
\label{fmeanmtm}
\end{center}
\end{figure*}
\section{Results}

\subsection{Particle Yields}
The mid-rapidity ($|y| <$ 0.1) BES data presented here are obtained using the STAR Time Projection chamber
(TPC) and Time-Of-Flight (TOF) detectors~\cite{tpctof}. 
The particles are identified by measuring the specific ionisation
energy loss in the TPC and the particle velocities using
TOF as a function of momentum. Figure ~\ref{fEngdndy} show the $dN/dy$
normalised to the average number of participant nucleus ($dN/dy/ \langle 0.5N_{part} \rangle $) vs. $\sqrt{s_{NN}}$ for $\pi^{\pm}$, $\emph{K}^{\pm}$, $\emph{p}$, and
$\bar{\emph{p}}$ for 0-5$\%$ centrality at BES energies and those from
published data at AGS~\cite{dataags1,dataags2,dataags3},
SPS~\cite{datasps}, RHIC~\cite{datastar}, and LHC energies~\cite{dataalice}. The errors on the points include
both statistical and systematic errors. The mid-rapidity yields of charged pions,
kaons, and anti-protons increases with
increasing collision energy whereas the protons yields decreases with
increase of collision energy. The results from BES data are in agreement with the general energy
dependence trend observed at AGS, SPS, top RHIC, and LHC energies. 

\subsection{Average Transverse Mass}
\begin{figure*}[!hbtb]
\begin{center}
\includegraphics[scale=0.42,height = 8.3cm,width =8.7cm]{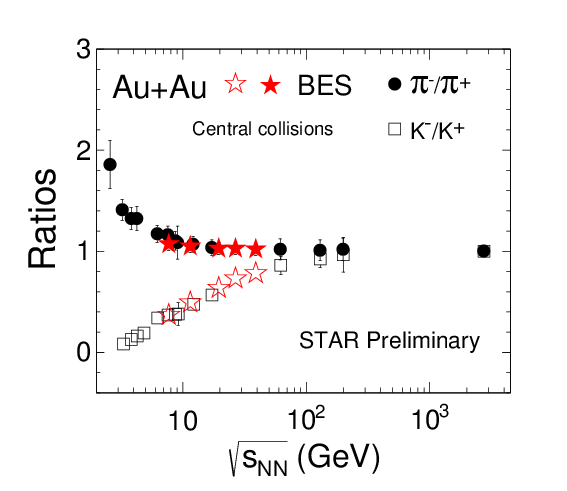}
\hspace{-0.64cm}
\includegraphics[scale=0.43,height = 8.3cm,width =8.7cm]{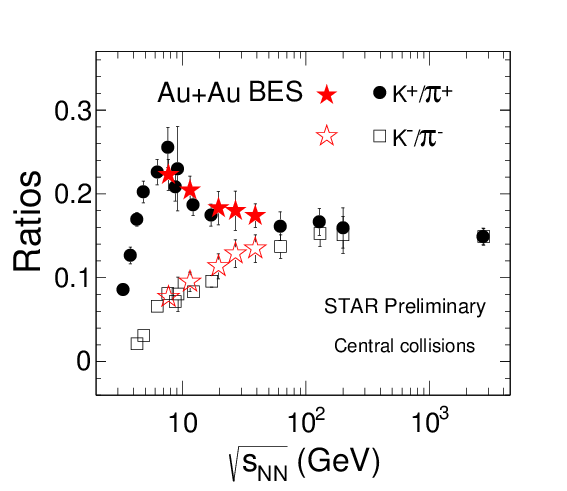}
\hspace{-0.65cm}
\includegraphics[scale=0.48]{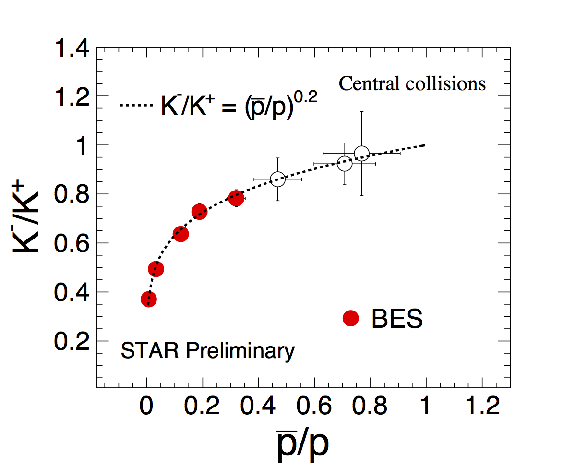}

\caption{Particle ratios as a function of center of mass energy in
  Au+Au collisions at BES energies. The results are compared
 with AGS~\cite{dataags1,dataags2,dataags3}, SPS~\cite{datasps}, RHIC~\cite{datastar},
 and LHC energies~\cite{dataalice}. The variation of ${\emph{K}^{-}}/\emph{K}^{+}$ as a function of
$\bar{\emph{p}}/\emph{p}$ for 0-5$\%$ centrality from SPS-LHC energies. Errors are statistical and
 systematic errors added in quadrature.}

\label{fEratios}
\end{center}
\end{figure*}
Figure~\ref{fmeanmtm} shows the energy dependence of $\langle m_{T}
\rangle -m $ for  $\pi^{\pm}$, $\emph{K}^{\pm}$, $\emph{p}$, and 
$\bar{\emph{p}}$ in central~(0-5 $\%$) Au+Au collisions at
BES energies along with data for 
Pb+Pb/Au+Au collisions from AGS~\cite{dataags1,dataags2,dataags3},
SPS~\cite{datasps}, RHIC~\cite{datastar}, and LHC
energies~\cite{dataalice}. 

The $\langle m_{T} \rangle -m $ increases with energy at lower energies, remains almost constant at
SPS and lower RHIC BES energies and then increases again towards higher energies up
to LHC energy. The behaviour of $ \langle m_{T} \rangle -m$ might
indicate the onset of the phase transition~\cite{meanmtqgp}. If the system is assumed to be 
thermodynamic, $\langle m_{T} \rangle -m $ and $\sqrt{s_{NN}} $ can be related to
temperature and entropy of the
system~($dN/dy \propto log \sqrt{s_{NN}}$), respectively. Based on this, the
constant value of $\langle m_{T} \rangle -m $ can be interpreted as a
signature of first order phase transition. $\langle m_{T} \rangle -m $
could be sensitive to several other effects~\cite{BMmeanmt}.
It is observed that with increasing centrality the values of $\langle
m_{T} \rangle$ increase, indicating that the development of stronger collective motion in more
central collisions and the difference between particles also
increases.

\subsection{Particle Ratios}

Figure~\ref{fEratios} shows the variation of different particle ratios as a function of center-of-mass energy in
  Au+Au collisions at BES energies and its comparison
  with the results from AGS~\cite{dataags1,dataags2,dataags3}, SPS~\cite{datasps}, RHIC~\cite{datastar},
 and LHC energies~\cite{dataalice}. The variation of ${\emph{K}^{-}}/\emph{K}^{+}$ as a function of
$\bar{\emph{p}}/\emph{p}$ for 0-5$\%$ centrality from SPS-LHC energies has been shown. Errors are statistical and
 systematic errors added in quadrature.\\
As the collision energy increases, $\pi^{-}/\pi^{+}$ ratio
decreases to unity whereas $\emph{K}^{-}/\emph{K}^{+}$ ratio rise systematically.
At higher energy, pair production, which results in the same number of
positive and negative pions 
dominates the resonance decays. Following this logic, the
$\pi^{-}/\pi^{+}$ ratio is supposed to reach unity as the energy goes up.
The $\emph{K}^{-}/\emph{K}^{+}$ ratio is indicative of the relative
contribution of associated and pair
production. The associated production mechanism can only produce
$\emph{K}^{+}$ via $N + N \to
N +X +K^{+}$, $\pi + N \to X + K^{+}$ where $N$ is nucleon and  $X$
is hyperon~($\Lambda$ or $\Xi$),
while the pair production mechanism produces $\emph{K}^{+}$and
$\emph{K}^{-}$ via $N + N \to N + N + \emph{K}^{+}+ \emph{K}^{-}$. 
The rise of $\emph{K}^{-}/\emph{K}^{+}$ ratio as a function of energy can be attributed to the nature
of kaon production channels. At lower energy the associated production dominates, due to a lower energy threshold. As the
energy increases, the pair production which produces the same number
of $\emph{K}^{+}$ and $\emph{K}^{-}$ becomes more significant. With increasing energy, the net baryon density
decreases and thus the associated production of $\emph{K}^{+}$
also decreases, while pair
production increases due to gluon-gluon fusion into strange
quark-antiquark pairs~\cite{pairprod1, pairprod2}. 
All these
results combined, when compared with previous experiments, seem to be consistent
with an enhancement in the strangeness production. At
lower energies due to the non-zero net baryon density in the
collisions zone, the
associated production of kaons with hyperons will be different from these produced
with anti-hyperons.

The $\emph{K}^{-}/\emph{K}^{+}$, which represents net-strange chemical
potential~($\mus$) vs. $\bar{\emph{p}}/\emph{p}$, representing net-baryon
chemical potential~($\mub$)
for 0-5$\%$ centrality in Au+Au collisions  together with results from
top RHIC, has been shown in Fig.~\ref{fEratios}. Both
ratios are affected by the net baryon content; they show a strong
correlation. In a hadron gas, both chemical potentials, $\mus$ and
$\mub$ depends on temperature and they follow the relation $\mus =
\mub/3$~\cite{musmub}. It is worth noting that at low energies, 
the absorption of antiprotons in the baryon-rich environment
plays a vital role.
\subsection{Chemical freeze-out}
At chemical freeze-out, inelastic collisions among the particles stop,
particle yields and ratios of particle yields get fixed.
Particle ratios are calculated taking
the ratios from the measured integrated yields. A set of different
particle ratios which involves the particle yields of $\pi^{\pm}$, $\emph{K}^{\pm}$, $\emph{p}$,
$\bar{\emph{p}}$, $\Lambda$, $\bar{\Lambda}$,
$\Xi^{-}$, $\bar{\Xi}^{+}$ can be
collectively used to extract the information on the chemical
freeze-out conditions. The extraction of freeze-out parameters is very senstive to 
the contribution of weak decays, commonly called feed-down. 
Experimentally different particles are corrected in
different ways. At STAR, proton yields have not
been corrected for feed-down contributions, and are commonly called
$``inclusive"$, where as $\pi$ and
$\Lambda$ yields have been corrected for the feed-down
from $K^{0}_{S}$, $\Xi$ and $\Xi^{0}$ weak decays, respectively. In
model, the appropriate feed-down as in experimental data has been considered. 
Different freeze-out
parameters are extracted using those ratios comparing with the
corresponding ratios calculated in the THERMUS model assuming chemical equilibrium.

\begin{figure*}[!hbtb]
\begin{center}
\includegraphics[scale=0.42,height = 6.8cm,width =7.3cm]{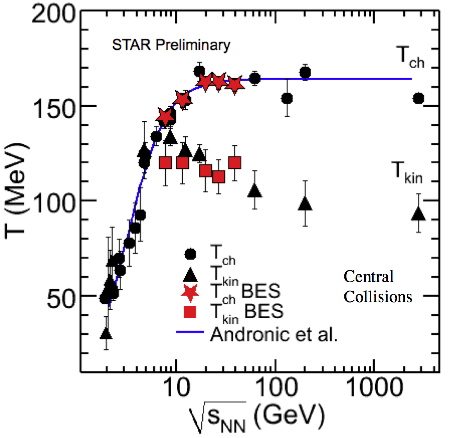}
\hspace{0.5cm}
\includegraphics[scale=0.42,height = 8.2cm,width=8.8cm]{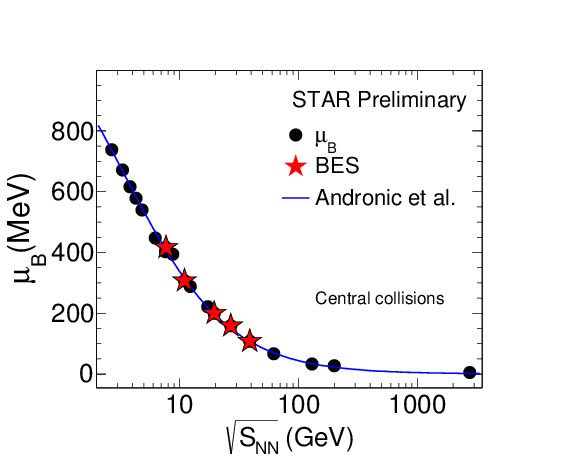}
\caption{The energy dependence of the extracted parameters
  $\tch$~(left) and $\mub$~(right) for central collisions. The curves
  represent the theoretical predictions~\cite{andro}.}
\label{ftmub}
\end{center}
\end{figure*}

In the THERMUS model, in thermodynamical equilibrium, the particle abundance of $i$-th particle~($N_{i}$) in a
system of volume $V$
can be given by~\cite{thms}
\begin{equation}
\frac{N_{i}}{V} = \frac{Tm_{i}^{2}g_{i}}{2\pi^{2}}\sum_{k=1}^{\infty}\frac{(\pm1)^{k+1}}{k} \exp(\beta k \mu_{i} )
 K_{2}\left( \frac{km_{i}}{T} \right) ,
\label{eqNV}
\end{equation}
where $T$ is the chemical freeze-out temperature, $m_{i}$ is
the particle mass, $g_{i}$ is the degeneracy, $\beta \equiv
\frac{1}{T}$, $K_{2}$ is the second order Bessel function 
and $\mu_{i}$ is the chemical potential of hadron species $i$ which is given by \\$\mu_{i} =
  B_{i}~\mu_{B} +S_{i}~\mu_{S}+Q_{i} ~\mu_{Q} $\\
where $B_{i}, S_{i}$ and $Q_{i} $ are the baryon number, strangeness and
charge, respectively, of species $i$ ~and ~$\mu_{B}$ , ~$\mu_{S}  $ and $\mu_{Q}
$ are the corresponding chemical potentials. In the model calculations
of particle ratios show a very good agreement with data at BES
energies studied for all centralities~\cite{sd}. The left plot 
in Fig.~\ref{ftmub} shows that the chemical freeze-out
temperature $\tch$ for central heavy-ion collisions as a function of energy and it increases with the 
increase of energy. The right plot of Fig.~\ref{ftmub} shows the
decrease of $\mub$ with increasing collision energy.
These measurements in the BES program have covered a wide range at
RHIC, from around $\mub = $ 20 MeV to about 420 MeV in the QCD phase diagram. The large value of $\mub$ at midrapidity
may indicate the formation of a high net-baryon density matter, which is expected to reach
a maximum value around 7.7 GeV~\cite{highmub}. The temperature of kinetic
freeze-out~($T_{kin}$), where elastic collisions stop and particle
spectra get fixed, obtained using a Blast-Wave fit to
the identified transverse momentum
spectra is found to takes place after chemical
freeze-out~\cite{lk2014}.
\subsection{Summary}
The identified particle production have been discussed in central Au+Au
collisions at all BES energies~($\sqrt{s_{NN}} = 7.7, 11.5, 19.6, 27, 39 $ GeV). The energy dependence of identified particle yields and 
ratios, the average transverse mass are discussed briefly. 
The measured particle ratios at mid-rapidity
have been used to extract the
parameters of chemical freeze-out, where the inelastic collisions of
hadrons have stopped. The chemical freeze-out parameters reflect 
the properties of the system at earlier stage compared with that at 
kinetic freeze-out.\\   
The BES energies along with the top RHIC energies 
have allowed one to access the region of the QCD phase diagram covering a
wide range of baryon
chemical potential ($\mu_B$) from 20 to 420 MeV corresponding to Au+Au collision energies
from $\sqrt {s_{NN} }=$ 200 down to 7.7 GeV, respectively. An
important $\tch, ~\mub$ point will be added at $\sqrt {s_{NN} }=$
14.5 GeV as a part of BES-I data which was taken in 2014 to cover a
large gap of $\mub$ between 11.5 and 19.6 GeV in phase diagram. 
Current lattice QCD calculations suggest that key features of the phase diagram
like the critical point and the first-order phase transition may lie within the $\mu_B$ reach of the
RHIC BES program~\cite{lhighmub}. In BES phase-II, a systematic measurement of the
yields of a variety of produced hadrons versus rapidity, centrality,
and beam energy will address various questions about the evolution of the
hadron yields between the initial hadronization and the final thermal
equilibrium~\cite{bes2} and about the possibility of successive
hadronization~\cite{bes3}. This could lead to further understanding
and refinement of the statistical models. In addition to BES program
at RHIC~\cite{netp14}, new experimental facilities have been
designed at the Facility for Antiproton and Ion Research (FAIR) at GSI and Nuclotron-based Ion Collider fAcility
(NICA) at JINR in order to search for the QCD critical
point~\cite{fair}.

\end{document}